\documentclass[aps,reprint,amsmath,amssymb,groupedaddress,notitlepage]{revtex4-1}
\usepackage[utf8]{inputenc}
\usepackage{bm}
\usepackage{enumerate}
\usepackage{graphicx}
\usepackage{cancel}
\usepackage{xcolor}
\usepackage{verbatim}
\usepackage{hyperref}
\hypersetup{colorlinks=true,
linkcolor=blue,
citecolor=blue,
urlcolor=blue,
filecolor=blue
}

\def\bra#1{\langle{#1}|}
\def\ket#1{|{#1}\rangle}
\newcommand{\bmk}[3]{\langle #1 | #2 | #3 \rangle}

\begin{abstract}
We propose an implementation of a singlet-only spin qubit in a GaAs-based triple quantum dot with a (1,4,1) charge occupation.
In the central multi-electron dot, the interplay between Coulomb interaction and an out-of-plane magnetic field creates an energy spectrum with a tunable singlet-triplet splitting, which can be exploited to create a six-particle singlet-only qubit with a qubit splitting that can straightforwardly be tuned over tens of $\mu$eV by adjusting the external magnetic field.
We confirm the full exchange-based electric control of the qubit and demonstrate its superior coherence properties due to its singlet-only nature.
\end{abstract}

\begin{document}

\title{Highly tunable exchange-only singlet-only qubit in a GaAs triple quantum dot}

\author{Arnau Sala}
\affiliation{Center for Quantum Spintronics, Department of Physics, Norwegian University of Science and Technology, NO-7491 Trondheim, Norway}

\author{Jørgen Holme Qvist}
\affiliation{Center for Quantum Spintronics, Department of Physics, Norwegian University of Science and Technology, NO-7491 Trondheim, Norway}

\author{Jeroen Danon}
\affiliation{Center for Quantum Spintronics, Department of Physics, Norwegian University of Science and Technology, NO-7491 Trondheim, Norway}

\date{\today}

\maketitle

Semiconductor spin qubits are among the most promising candidates for the physical realization of quantum processors~\cite{Hanson2007,Zwanenburg2013}.
Multi-spin exchange-only (XO) qubits, in particular, have drawn much attention in recent years since they offer fast qubit manipulation and full electric control~\cite{Russ2017,Laird2010,Gaudreau2011,Medford2013,Medford2013a,Eng2015,Andrews2019}.
However, rapid decoherence of the qubit---due to magnetic noise from randomly fluctuating nuclear spins~\cite{Hung2014,Peterfalvi2017}, electric noise in the qubit's environment~\cite{Russ2015,Martins2016,Yoneda2018}, electron-phonon coupling~\cite{Taylor2013,Zhang2018,Sala2018}, and other spin-mixing mechanisms~\cite{Golovach2004,Raith2012,Raith2012a,Hofmann2017}---still causes the usable operation time of most XO qubits to be too short for scaling up.
Besides, the typically small qubit splitting~\cite{Medford2013a,Russ2017} hinders the long-distance coupling of XO qubits via, e.g, microwave resonators, where a large qubit splitting is required for fast two-qubit gates~\cite{Srinivasa2016,Landig2018,Harvey2018}.

There have been several proposals put forward to increase the coherence time of quantum-dot-based XO qubits while retaining their conceptual simplicity and ease of manipulation.
Of special interest are (\textit{i}) proposals to suppress the effects of charge noise and electron-phonon interaction, via a symmetric operation of the qubit or operating at a sweet spot (SS)~\cite{Reed2016,Shim2016,Malinowski2017,Zhang2018}, and (\textit{ii}) proposals to reduce magnetic noise or suppress its effects, either by isotope purification or by constructing decoherence-free qubit subspaces~\cite{Bacon2000,Eng2015,Friesen2017,Sala2017,Russ2018}.

In the exchange-only singlet-only (XOSO) spin qubit proposed in Ref.~\cite{Sala2017}, the leading effects of magnetic noise are suppressed by encoding the qubit states in a four-electron singlet-only subspace, while electric noise can be mitigated by operating the system symmetrically at a SS.
However, the exceptionally long coherence time of the qubit comes at the cost of an increase in device complexity (a quadruple quantum dot in a T-geometry) and the proposal suffers from the common problem with XO qubits of having a relatively small qubit splitting.

Here, we propose a GaAs-based implementation of the XOSO qubit that overcomes both drawbacks and, furthermore, has a qubit splitting that is straightforwardly tunable over a large range of energies.
The reason why the XOSO qubit of Ref.~\cite{Sala2017} used a fourth quantum dot is that the qubit splitting scales with the singlet-triplet splitting of the ``central'' two electrons:
Implementing the same qubit in a linear triple dot in a (1,2,1) charge configuration is in principle possible, but results in a qubit with a splitting of the order of the orbital level splitting on the central dot ($\sim$~meV), which is too large for practical purposes.
In Ref.~\cite{Russ2018} it was pointed out that one can implement the same qubit in a Si-based triple dot, where the on-site singlet-triplet splitting is typically set by the valley splitting, which can be 20--200~$\mu$eV.
The drawback of this proposal is that (\textit{i}) the magnitude of the valley splitting is hard to control or predict in practice~\cite{Zwanenburg2013} and (\textit{ii}) uncontrollable phase differences between valley couplings on different dots can severely affect the exchange effects used to define and operate the qubit~\cite{Culcer2010a}.
Besides, Si can be purified to be almost nuclear-spin-free, which eradicates the need for a singlet-only qubit~\cite{Eng2015}.

The solution is to tune the triple quantum dot to a (1,4,1) charge configuration and apply an out-of-plane magnetic field.
On the central dot, the interplay between the magnetic field and the Coulomb interaction between the electrons results in an energy spectrum with many crossings between levels with different total spin and orbital angular momentum.
For the case of four electrons, the ground state changes from a triplet to a singlet character, typically at a moderate field of $\sim 100$~mT~\cite{Eto1997}.
Tuning close to this crossing and adding the singly-occupied outer dots to the picture yields a XOSO qubit where the singlet-triplet splitting on the central dot, and thus the qubit splitting, can be tuned by adjusting the external magnetic field.
This yields a superior GaAs-based XOSO qubit that can be hosted in a simple linear array of three dots and has a qubit splitting that is straightforwardly tunable from zero to tens of $\mu$eV.

\textit{Multi-electron dot.}---The single-particle Hamiltonian of an electron labeled $i$ in a two-dimensional planar quantum dot, assuming a parabolic confinement and an external magnetic field perpendicular to the plane, is
\begin{align}\label{eq:ham0}
H^{(i)}_0 = {}&{}  \frac{[\mathbf p_i + e \mathbf A (\mathbf r_i)]^2}{2 m^*}  + \frac{1}{2} m^* \omega_0^2 r_i^2 +  \frac{1}{2}g \mu_{\rm B} B \sigma^{z}_{i},
\end{align}
where $\mathbf A (\mathbf r) = \frac{1}{2}B(x\hat y - y\hat x)$  is the vector potential, $\omega_0$ sets the effective radius of the dot in the absence of a magnetic field $\sigma_0=\sqrt{\hbar/m^* \omega_0}$, $g$ is the $g$-factor of the host material, and $\sigma^{z}$ is the third Pauli matrix.
The eigenstates of this Hamiltonian are the Fock-Darwin states,
\begin{align}\label{eq:fock-darwin}
\psi_{n,l,\eta} (\mathbf r) = {}&{} \sqrt{\frac{n!}{\pi \sigma^2(n+|l|)!}} \rho^{|l|} e^{-\rho^2/2} L_n^{|l|}(\rho^2) e^{-i l \theta},
\end{align}
in terms of the dimensionless polar coordinates $\rho=r/\sigma$ and $\theta$.
We used $\sigma=\sqrt{\hbar/m^* \Omega}$, with $\Omega = \sqrt{\omega_0^2 + \omega_c^2/4}$ and $\omega_c = e B/m^*$, and $L_a^{b}(x)$ is the associated Laguerre polynomial.
The quantum numbers $n \in \mathbb{N}_0$, $l \in \mathbb{Z}$, and $\eta = \pm 1$ label the radial state, orbital angular momentum, and spin of the electron, respectively.
The corresponding eigenenergies are (we will set $\hbar = 1$ from now on)
\begin{align}\label{eq:FD}
E_{n,l,\eta} = {}&{} \Omega (2 n + |l| + 1) - \frac{1}{2} \omega_c l + \frac{1}{4}g\omega_c\frac{m^*}{m_e}\eta.
\end{align}

In order to find the approximate eigenenergies and spin structure of multi-electron states in the presence of electron-electron interactions, we follow the method used in~\cite{Eto1997,Rontani1999}, see the Supplemental Material in~\cite{SeeSM} for the details.
We create a many-particle basis of antisymmetrized products of single-particle states (\ref{eq:fock-darwin}), where we restrict ourselves to the states with $n \leq 1$ and $|l| \leq 3$, which corresponds to including all single-particle levels up to $\sim 4 \Omega$ at small fields.
In the thusly constructed basis we evaluate all matrix elements of the interaction Hamiltonian
\begin{align}
	V = \sum_{i<j} \frac{e^2}{4 \pi \varepsilon | \mathbf r_i - \mathbf r_j |},
\end{align}
and the eigenstates and -energies of the full many-particle Hamiltonian $H_1 = \sum_{i}   H_0^{(i)} + V$ can then be found from numerical diagonalization or, in the weak-interaction limit characterized by $\kappa \equiv e^2/4\pi\varepsilon\sigma_0\omega_0 \ll 1$, from perturbation theory in $\kappa$.
For few particles and not too large $\kappa$ (we consider up to five electrons and $\kappa \leq 1.5$) the low-energy part of the spectrum of $H_1$ will resemble the exact many-particle spectrum fairly accurately~\cite{Eto1997,Reimann2002}.

\begin{figure}
\includegraphics[width=\linewidth]{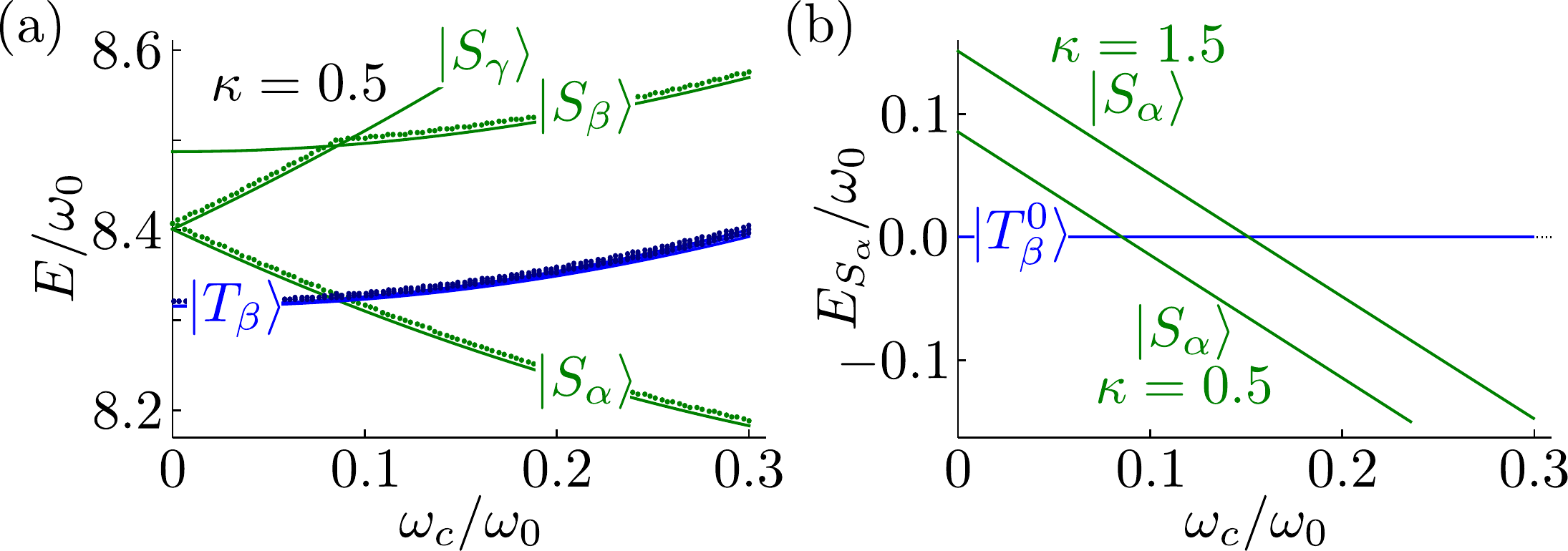}
\caption{(a) Field-dependent low-energy part of the spectrum of a four-electron quantum dot with $\kappa = 0.5$ and $g=-0.4$. Dots present numerical results and solid lines the perturbative results of (\ref{eq:eigs}).
(b) The numerically evaluated energy of the state $\ket{S_\alpha}$ (green lines) relative to $\ket{T_\beta^0}$ for two values of $\kappa$.\label{fig:fig1}}
\end{figure}

In Fig.~\ref{fig:fig1}(a) we present typical results for the lowest few levels for the case of four electrons, where we set $\kappa = 0.5$ and $g=-0.4$.
The dots show the numerically calculated lowest five eigenenergies, where green(blue) dots indicate a state with a four-particle spin singlet(triplet) structure.
The three triplet states are labeled $\ket{T_\beta}$ and have the largest weight in the orbital configuration $(0,0)^2(0,1)^1(0,-1)^1$, where $(n,l)^m$ means $m$ electrons in the orbital state $(n,l)$~\cite{Eto1997}.
The three lowest singlet states, labeled $\ket{S_{\alpha,\beta,\gamma}}$, live mostly in the orbital configurations $(0,0)^2(0,1)^2$, $(0,0)^2(0,1)^1(0,-1)^1$, and $(0,0)^2(0,-1)^2$, respectively.

For small $\kappa$, these lowest eigenenergies can also be approximated through perturbation theory in the interaction Hamiltonian $V$.
Up to second order in $\kappa$ this yields for the lowest six states the generic expression
\begin{align}\label{eq:eigs}
	E_\nu = 6 \Omega -\frac{L}{2}\omega_c + \frac{S}{2}g\omega_c\frac{m^*}{m_e} + c^{(\nu)}_1 \!\kappa \sqrt{\Omega\omega_0} + c^{(\nu)}_2\! \kappa^2\omega_0,
\end{align}
where $L$ and $S$ denote the total orbital and spin angular momentum along $\hat z$ of the four electrons.
The coefficients $c^{(\nu)}_{1,2} \sim 1$ differ per state $\ket{\nu}$ but can be found explicitly, see~\cite{SeeSM} for their exact values.
The resulting energies $E_\nu$ are plotted in Fig.~\ref{fig:fig1}(a) as solid lines, and show good agreement with the numerics.
For larger $\kappa$ the perturbation theory breaks down, but the low-energy part of the spectrum is qualitatively the same.
This suggests that one can use Eq.~(\ref{eq:eigs}) to describe the $E_\nu$ if one treats the coefficients $c_{1,2}^{(\nu)}$ as fit parameters to the numerical data.
As illustrated in~\cite{SeeSM} for the case $\kappa=1.5$, this still leads to excellent agreement.
In Fig.~\ref{fig:fig1}(b) we show the numerically evaluated energy of the state $\ket{S_\alpha}$ relative to $\ket{T_\beta^0}$ as a function of $\omega_c$, for $\kappa=0.5$ and $\kappa=1.5$.
In both cases the splitting between $\ket{S_\alpha}$ and $\ket{T_\beta^0}$ is to good approximation linear in $\omega_c$ in the regime of interest, and the ground state changes from a spin triplet to a singlet around $\omega_c/\omega_0 \sim 0.1$.
These two generic features are the key ingredients for our qubit proposal.

\begin{figure}
	\includegraphics[width=\linewidth]{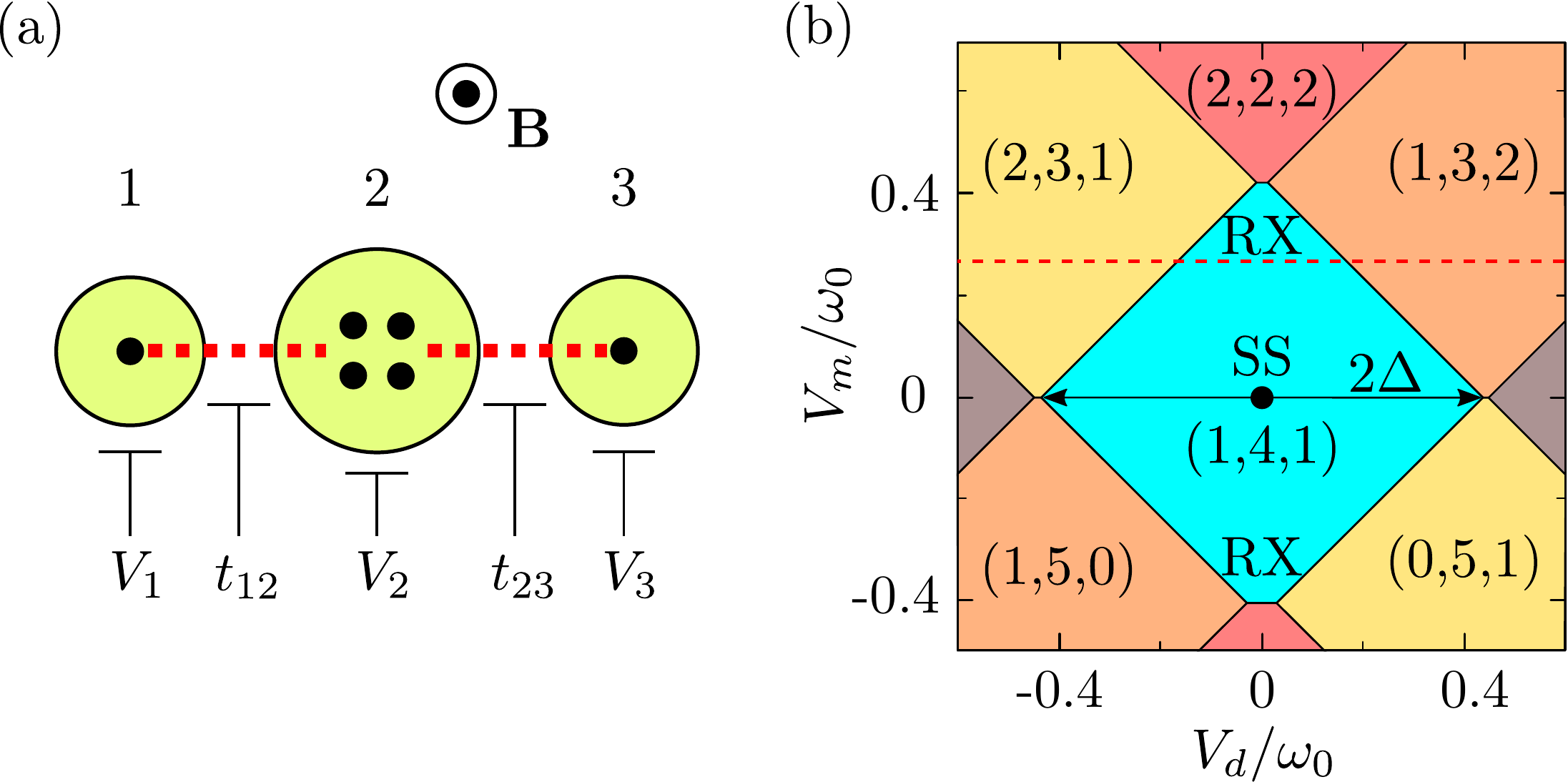}
	\caption{(a) Sketch of the linear triple-dot setup in a (1,4,1) charge configuration with a perpendicular magnetic field applied.
		(b) Six-electron charge stability diagram around the (1,4,1) ground state, as a function of $V_m$ and $V_d$.\label{fig:fig2}}
\end{figure}

\textit{Triple-dot six-electron states.}---We will construct our qubit in two \emph{six}-electron states hosted in a linear arrangement of three quantum dots with a perpendicular magnetic field applied, such as sketched in Fig.~\ref{fig:fig2}(a), where
the effective on-site potentials $V_i$ and the interdot tunnel couplings $t_{ij}$ can be controlled through nearby gate electrodes, as schematically indicated.
We describe this system using a simple Hubbard-like Hamiltonian~\cite{Burkard1999,Taylor2013,Sala2017},
\begin{align}\label{eq:hubbard}
    {H} = {} & {} \sum_{i=1}^3\left( H_1^{(i)} - V_i{n}_i\right) + \sum_{\langle i,j\rangle}U_c{n}_i{n}_j
    -\!\! \sum_{\langle i,j\rangle,\eta}\! \frac{t_{ij}}{\sqrt{2}}{c}_{i\eta}^\dagger{c}_{j\eta},
\end{align}
where ${n}_i=\sum_\eta{c}_{i\eta}^\dagger{c}_{i\eta}$ is the number operator for dot~$i$, ${c}_{i\eta}$ annihilates an electron on dot $i$ with spin $\eta$, $U_c$ accounts for the cross-capacitance between neighboring dots, and $H_1^{(i)}$ is the single-dot many-particle Hamiltonian for dot $i$ as described above.
We thus made several simplifying assumptions:
(\textit{i}) The gate-induced potentials are smooth enough so that they affect all electronic orbitals in the same way.
(\textit{ii}) The separation between the dots is large enough to allow us to treat the \emph{inter}dot electrostatic energy as being dependent only on the $n_i$ and not on the exact orbital configuration of the electrons on the neighboring dots.
(\textit{iii}) All tunneling processes we will consider below mostly involve a (0,0)-orbital on a lateral dot and a $(0,\pm 1)$-orbital on the central dot; since all $(0,\pm 1)$-orbitals have the same radial structure we assume that this allows us to use tunneling coefficients $t_{ij}$ that are independent of the exact electronic orbitals involved.

We first study the electrostatic properties of ${H}$ by diagonalizing the first two terms in Eq.~(\ref{eq:hubbard}). The charge stability diagram in Fig.~\ref{fig:fig2}(b) shows the resulting six-electron ground-state charge configuration ($n_1$, $n_2$, $n_3$), where $n_i$ is the number of electrons on dot $i$, as a function of the detuning parameters $V_d=\frac{1}{2}\left(V_3-V_1\right)$ and $V_m=\frac{1}{2}\left(V_1+V_3\right)-V_2$.
We fixed $V_1 + 4V_2+V_3$ and focused on the regime around the (1,4,1) state.
As indicated in Fig.~\ref{fig:fig2}(a), we assumed different dot sizes, $\sigma_0=30$~nm for the central dot and $\sigma_0=20$~nm for the lateral dots, which results in a good ratio between the orbital splitting on the outer dots and the splitting of the many-electron states in the middle dot~\cite{footnote1}.
Furthermore, we used  $U_c=0.2\, \omega_0$ (where $\omega_0$ is the bare level splitting on the \emph{central} dot) and set $\omega_{c}/\omega_0 =0.1$, $\kappa = 0.5$, and $m^*/m_e = 0.067$.

In the (1,4,1) region the four lowest-energy six-particle states with ${S}^2 = 0$ can be written as
\begin{align}
    \ket 0 = {}&{} \ket{S_{\alpha} S_{(13)}}, 
    \label{eq:ket0}
    \\
    \ket 1 = {}&{} \frac{1}{\sqrt 3}\left[\ket{T_{\beta}^{0} T_{(13)}^{0}} - \ket{T_{\beta}^{-} T_{(13)}^{+}} - \ket{T_{\beta}^{+} T_{(13)}^{-}} \right],
    \label{eq:ket1}
    \\
    \ket 2 = {}&{} \ket{S_{\beta} S_{(13)}}, 
    \label{eq:ket2}
    \\
    \ket 3 = {}&{} \ket{S_{\gamma} S_{(13)}}, 
    \label{eq:ket3}
\end{align}
where $\ket{S_{(13)}}$ and $\ket{T_{(13)}}$ indicate pairing in a singlet or triplet state of the two electrons in the outer dots, and $\ket{S_{\alpha,\beta,\gamma}}$ and $\ket{T_{\beta}}$ are the lowest four-particle singlets and triplet on the central dot, see above.

\textit{The qubit.}---We propose to tune close to the degeneracy of $\ket{S_\alpha}$ and $\ket{T_\beta}$ on the central dot, which for $\sigma_0 = 30$~nm happens at $B \approx 75~$mT.
The two lowest-energy singlet states $\ket 0$ and $\ket 1$ can then be used as qubit basis, and the singlets $\ket 2$ and $\ket 3$ will be split off by an energy much larger than the qubit splitting.

We assume that $t/\Delta \ll 1$, with $t$ the magnitude of the tunnel couplings (typically $t \sim 10~\mu$eV) and $2\Delta$ the width of the (1,4,1) region, see its definition in Fig.~\ref{fig:fig2}(b).
Then we can treat the tunnel coupling perturbatively for most of the (1,4,1) region, and we thus project the full Hamiltonian (\ref{eq:hubbard}) onto the qubit subspace by means of a Schrieffer-Wolff transformation~\cite{SeeSM}, yielding to order $t^2$
\begin{equation}
    {H}_\text{qb} = \frac{1}{2}\left(E_{ST} + J_z\right)\sigma_z + J_x\sigma_x,
    \label{eq:qubit_Hamiltonian}
\end{equation}
where $\sigma_{x,z}$ are Pauli matrices.
The qubit splitting is dominated by the singlet-triplet splitting on the central dot $E_{ST}=E_{T^0_\beta} - E_{S_\alpha}$ [see Fig.~\ref{fig:fig1}(b)], which follows to good approximation from the expressions given in (\ref{eq:eigs}),
\begin{align}
	E_{ST} \approx {}&{}  \gamma_0\omega_0 + \omega_c,
\end{align}
with $\gamma_0 = -0.235\,\kappa + 0.128\,\kappa^2$, accurate for $\kappa \lesssim 0.5$ (see \cite{SeeSM} for all derivations and an explicit expression for $\gamma_0$).
We wrote $E_{ST}$ here up to linear order in $\omega_c/\omega_0$; the next correction is smaller by a factor $\sim 10^{-2} \kappa \omega_c/\omega_0$.
We emphasize that through $\omega_c \propto B$ this term, and thus the qubit splitting, can be easily tuned over tens of $\mu$eV.

Close to the line where $V_d=0$ and assuming approximately symmetric tunnel couplings $t_{12}\approx t_{23}$, the two exchange terms read as~\cite{SeeSM}
\begin{align}
J_z \approx {} & {} -t^2\left[ \frac{\Delta}{\Delta^2 - V_m^2} + \frac{3(\Delta + \omega_c)}{(\Delta + \omega_c)^2 - V_m^2} \right], \\
J_x \approx {} & {} \frac{\sqrt 6 t \Delta}{\Delta^2 - V_m^2} \left[ \delta t + \frac{2 t V_m}{\Delta^2 - V_m^2}V_d \right], \label{eq:jx}
\end{align}
for $\Delta$ as defined in Fig.~\ref{fig:fig2}(b), and with $t = \frac{1}{2}(t_{12} + t_{23})$ and $\delta t = t_{12}-t_{23}$.
We see that $J_z$ in general presents a small tuning-dependent correction to the qubit splitting, which is dominated by $E_{ST}$, whereas $J_x$ provides a coupling to $\sigma_x$ linear in $\delta t$ and/or $V_d$ (depending on tuning), which can be used to drive Rabi oscillations.

\begin{figure}
	\includegraphics[width=\linewidth]{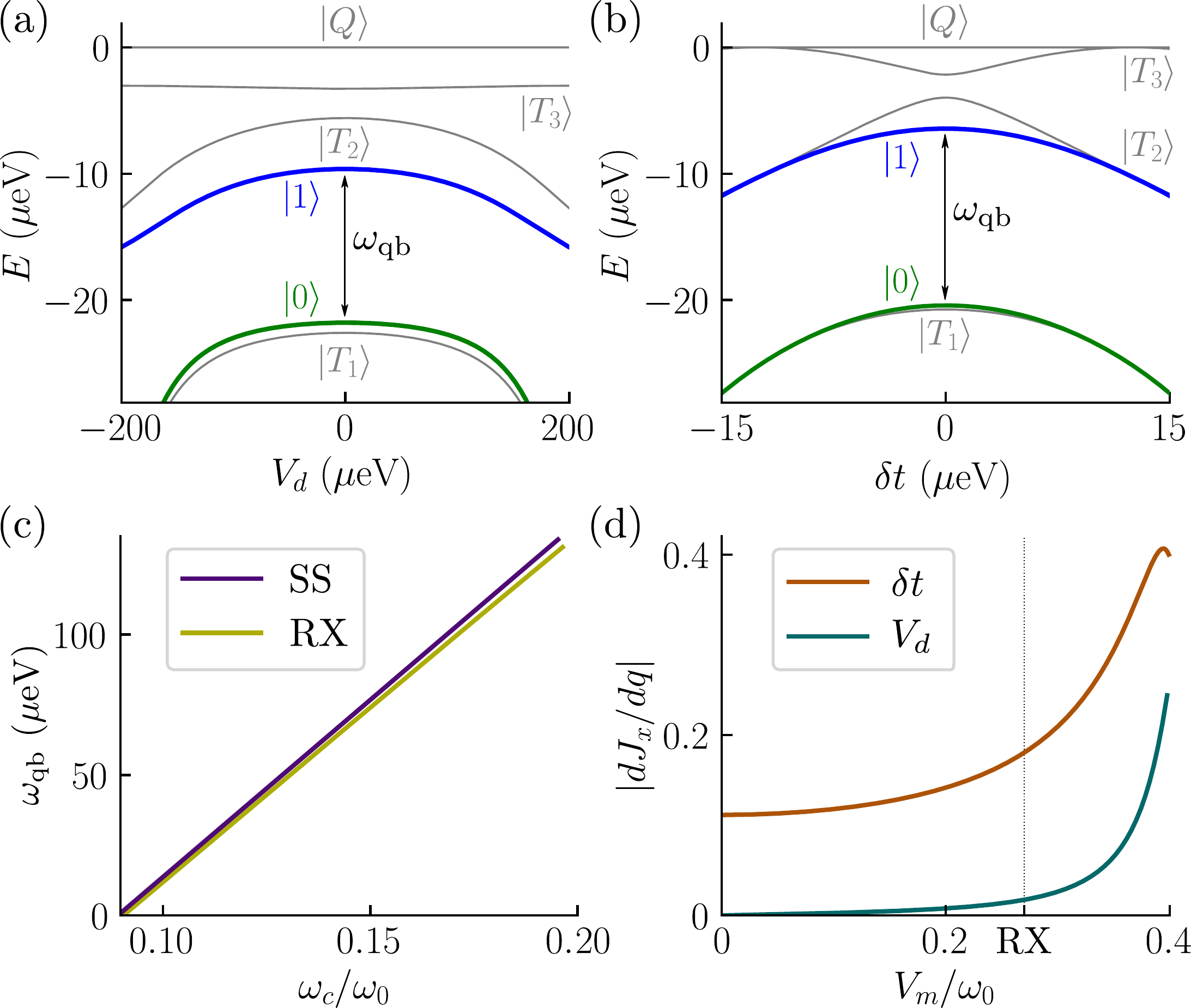}
	\caption{Low-energy part of the spectrum of the Hamiltonian (\ref{eq:hubbard}) (a) as a function of $V_d$ at $V_m/\omega_0 = 0.27$ and (b) as a function of $\delta t$ at the SS, $V_d = V_m = 0$. The green and blue lines show the spin-singlet qubit states $\ket 0$ and $\ket 1$ respectively; the grey lines show the spin triplet and quintuplet states.
	(c) The qubit splitting as a function of the magnetic field, where $\omega_c / \omega_0 = 0.1$ corresponds to $B \approx 75$~mT.
	(d) The derivative $dJ_x/dq$ for $q \in \{ \delta t, V_d\}$ as a function of $V_m$ and at $V_{d} = 0$.}
	\label{fig:specH}
\end{figure}

We now discuss two regimes of special interest in the charge stability diagram shown in Fig.~\ref{fig:fig2}(b):
(\textit{i})
In the resonant-exchange (RX) regime, close to the top and bottom of the (1,4,1) region, the strong coupling to the other charge states offers fast qubit control through $V_d$~\cite{Medford2013a}.
In Fig.~\ref{fig:specH}(a) we show the lowest-lying states as a function of $V_d$ along the horizontal dashed line in Fig.~\ref{fig:fig2}(b) ($V_m/\omega_0 = 0.27$) calculated from the Hamiltonian as given in (\ref{eq:hubbard}), where we ignored the Zeeman splitting for clarity.
We used the same parameters as in Fig.~\ref{fig:fig2}(b) and further set $t= 25~\mu$eV and $\delta t = 0$.
We labeled the two qubit states $\ket 0$ and $\ket 1$, three spin triplets $\ket{T_{1,2,3}}$, and a spin quintuplet $\ket{Q}$; including the Zeeman effect, a triplet(quintuplet) acquires an additional threefold(fivefold) splitting of 1.7~$\mu$eV for $\omega_c/\omega_0 = 0.1$.
(\textit{ii})
In the center of the (1,4,1) region we find a SS where the qubit is to linear order insensitive to fluctuations of the potentials $V_i$, offering some protection against charge noise.
In Fig.~\ref{fig:specH}(b) we show the spectrum at the SS for the same parameters as in (a), now as a function of $\delta t$ while setting $V_d = 0$.
At the SS exchange effects are much smaller and thus the qubit splitting is closer to $E_{ST}$ ($\approx 18.3~\mu$eV for $\omega_c/\omega_0 = 0.1$), but apart from that the spectrum looks similar to the RX regime.

In Fig.~\ref{fig:specH}(c) we plot the qubit splitting $\omega_{\rm qb}$ as a function of the magnetic field, in the RX regime ($V_m = 0.27$, $V_d = 0$, yellow line) and at the SS (purple line).
This confirms the high degree of tunability of our qubit.
We further note how the spectra in Fig.~\ref{fig:specH}(a,b) strongly resemble those in the XOSO spin-qubit proposals of Refs.~\cite{Sala2017,Russ2018}, the main difference being the large and straightforwardly tunable qubit splitting $\omega_{\rm qb} \propto B$ in our proposal.
This permits an efficient and adaptable coupling to other systems such as microwave cavities which can be used to couple distant qubits~\cite{Burkard2019}.

\textit{Qubit operation.}---Single-qubit rotations can be performed via resonant Rabi driving, using a sinusoidal modulation of a tuning parameter $q=\{ V_d, V_m, t, \delta t \}$ with a small amplitude $\tilde q$ and frequency $\omega$, i.e., $q(t) = q_0 + \tilde q\, \sin(\omega t)$. For small enough $\tilde q$ the qubit Hamiltonian (\ref{eq:qubit_Hamiltonian}) can be approximated as
\begin{equation}
    {H}_\text{qb} 
    = \frac{1}{2} \omega_{\rm qb} \sigma_z + A_q \sin ( \omega t ) \sigma_x,
\end{equation}
where $A_q = \tilde q\, (dJ_x/dq)_{q = q_0}$.
Driving the qubit resonantly, $\omega = \omega_{\rm qb}$, then induces Rabi oscillations with a frequency $A_q$.
At the RX regime, where we can use $V_d$ as the driving parameter, an amplitude of $\tilde V_{d}= 5$--$10~\mu$eV gives a Rabi period of $T_{\text{Rabi}}\approx 20$--$40$~ns.
At the SS Rabi rotations are much more efficient via a driving of $\delta t$, which gives a period of $T_{\text{Rabi}}\approx 20$~ns for an amplitude $\delta \tilde t = 2~\mu$eV. Fast qubit rotations can therefore be achieved both in the RX regime and at the SS.
In Fig.~\ref{fig:specH}(d) we plot the ``efficiency'' $dJ_x/dq$ of the two driving parameters $q \in \{ \delta t,V_d\}$ as a function of $V_m$, along the line $V_{d,0} = 0$.
We see that at the SS the sensitivity to $V_d$ vanishes, in accordance with Eq.~(\ref{eq:jx}), whereas driving of $\delta t$ stays effective all the way down to $V_m = 0$.

Qubit initialization and readout can be accomplished by standard spin-to-charge conversion, i.e., pulsing the qubit to one of the neighboring charge configurations that has only one low-lying six-particle singlet state.
For example, when tuning into the (1,3,2)/(2,3,1) charge regions, only the qubit state $\ket{0}$ is adiabatically connected to the new ground state charge configuration.
This allows for initialization in $\ket 0$ as well as read-out of the qubit by means of charge detection.

\textit{Decoherence.}---In most GaAs-based spin qubits the main source of decoherence is the fluctuating bath of nuclear spins that couples to the electron spins via hyperfine interaction.
On a mean-field level, the effect of this interaction can be described by the Hamiltonian $H_\text{hf} = \frac{1}{2} g \mu_{\rm B}\sum_{i} \mathbf K_i\cdot \boldsymbol{\sigma}_{i}$, with $\mathbf K_i$ a random effective nuclear field acting on electron $i$, typically of the order or a few mT.
In the device we propose in this paper, both qubit states are singlets and therefore the qubit splitting is not directly influenced by any intrinsic or external (gradient) of Zeeman fields acting on the electrons, thereby reducing the hyperfine-induced decoherence dramatically~\cite{Sala2017,Russ2018}.
The dominating remaining effect of the nuclear fields is the coupling of the qubit states to nearby triplet states, which leads to random higher-order shifts of the qubit levels~\cite{Sala2017}.
The time scale of this residual hyperfine-induced dephasing can be estimated as $T_2^* \sim A_q\hbar(\delta \varepsilon)^2/\sigma_{K}^4$, where $\delta\varepsilon$ is the energy splitting between $\ket 0$ and $\ket{T_1}$, see Fig.~\ref{fig:specH}(a,b)~\cite{SeeSM}.
For the parameters considered here, we find $T_2^* \sim 0.5$--$5~\mu$s~\cite{SeeSM}.

Another source of decoherence for exchange-based qubits are low-frequency fluctuations in the electrostatic environment of the system.
A common way to mitigate such charge noise is to operate the qubit at the SS, where the qubit splitting is insensitive to fluctuations in the potentials $V_i$ to leading order.
Away from the SS, the effects of charge noise enter the qubit splitting through exchange coupling [$J_z$ in (\ref{eq:qubit_Hamiltonian})]; but since both $E_{ST}$ and $J_z$ can be tuned in experiment, a great advantage of our proposal is that the relative contribution of $J_z$ to $\omega_{\rm qb}$ can in principle be made as small as desired.

Finally, qubit relaxation via electron-phonon coupling causes qubit decoherence.
The relaxation rate can be estimated  using Fermi's golden rule and depends on the qubit splitting and on the strength of the exchange interaction~\cite{Sala2017}.
In the RX regime, where the qubit splitting can be extensively tuned through $\omega_c$, we estimate relaxation rates from $\Gamma_\text{rel}\sim 1~$GHz for $\omega_{\rm qb} \sim 50~\mu$eV to $\Gamma_\text{rel}\sim 1~$MHz for $\omega_{\rm qb} \sim 10~\mu$eV.
And, as is common in exchange-based qubits~\cite{Sala2017,Sala2018}, the relaxation rate is strongly suppressed as we approach the SS.

\textit{Conclusions.}---We propose a six-electron exchange-only singlet-only spin qubit hosted in a GaAs linear triple quantum dot.
Its singlet-only nature makes the qubit intrinsically insensitive to randomly fluctuating nuclear fields.
The qubit can be operated fully electrically, either in an RX regime which enables fast qubit operations, or at a SS where the qubit is better protected against charge noise.
Furthermore, the fact that the qubit splitting is highly tunable over a large range of energies allows for efficient and adaptable coupling to microwave resonators, enabling coupling of distant qubits.

\textit{Acknowledgements.}---This work is part of FRIPRO-project 274853, which is funded by the Research Council of Norway (RCN), and was also partly supported by the Centers of Excellence funding scheme of the RCN, project number 262633, QuSpin.

%


\clearpage
\onecolumngrid
\begin{center}
\textbf{\large Highly tunable exchange-only singlet-only qubit in a GaAs triple quantum dot:\\ Supplemental Material}
\end{center}

\vspace{1ex}
\begin{center}
Arnau Sala, Jørgen Holme Qvist, and Jeroen Danon\\
\textit{\small Center for Quantum Spintronics, Department of Physics,\\ Norwegian University of Science and Technology, NO-7491 Trondheim, Norway}
\end{center}
\normalfont

\setcounter{equation}{0}
\setcounter{section}{0}
\setcounter{figure}{0}
\setcounter{page}{1}
\makeatletter



\renewcommand{\theequation}{S\arabic{equation}}
\renewcommand{\thefigure}{S\arabic{figure}}

\vspace{5ex}

\begin{NoHyper}

In this supplemental material we complement the results presented in the main text with several more detailed discussions. We included (\textit{i}) a detailed explanation of how to construct the many-electron Hamiltonian and the derivation of analytical approximate expressions for the energy spectrum of a multi-electron quantum dot; (\textit{ii}) the derivation of the qubit Hamiltonian via a Schrieffer-Wolff transformation; and (\textit{iii}) an estimate of the residual effects of the hyperfine coupling to the fluctuating nuclear spin baths on the qubit's coherence properties.

\vspace{4ex}

\section*{Electronic states in a multi-electron quantum dot}

\subsection*{Single-particle states in the presence of a perpendicular magnetic field}

In this section we investigate the spectrum of a multi-electron quantum dot in the presence of a magnetic field perpendicular to the plane of the dot, where we will largely follow the method used in Ref.~\onlinecite{SEto1997}. We assume strong confinement of the electrons along the $z$-direction, perpendicular to the plane, and a circularly symmetric parabolic in-plane  confinement. Under these assumptions we write a single-particle Hamiltonian in the $xy$-plane for electron $i$:
\begin{align}\label{s:ham0}
H_0^{(i)} = {}&{}  \frac{1}{2 m^*} [\mathbf p_i + e \mathbf A (\mathbf r_i)]^2 + \frac{1}{2} m^* \omega_0^2 \mathbf r_i^2 + \frac{1}{2}g\mu_{\rm B}B \sigma_i^{z},
\end{align}
where $\mathbf A (\mathbf r) = \frac{1}{2}B(x\hat y - y\hat x)$ is the vector potential describing the magnetic field ${\bf B} = B\hat z$, $m^*$ is the effective mass of the electrons, $\omega_0$ defines the strength of the in-plane confinement, such that $\sigma_0=\sqrt{\hbar/(m^* \omega_0)}$ gives the effective radius of the dot in the absence of a magnetic field, and $g$ is the effective $g$-factor of the host material.

The eigenstates of this Hamiltonian are the Fock-Darwin states,
\begin{align}
\psi_{n,l,\eta} (\mathbf r_i) = {}&{} \sqrt{\frac{n!}{\pi \sigma^2 (n+|l|)!}}
\, \rho_i^{|l|} e^{-\rho_i^2/2} L_n^{|l|}\left( \rho_i^2 \right) e^{-i l \theta_i},\label{s:fdstates}
\end{align}
written in terms of polar coordinates $\rho_i = r_i/\sigma$ and $\theta_i$.
Here, $\sigma = \sqrt{\hbar/m^* \Omega}$ is the magnetic-field-dependent effective dot radius, with $\Omega = \sqrt{\omega_0^2 + \omega_c^2/4}$ and $\omega_c = e B/m^*$, and
$L_a^{b}\left( x \right)$ is the associated Laguerre polynomial.
The quantum number $n = 0,1,2,\dots$ labels the radial orbital degree of freedom, the quantum number $l \in \mathbb{Z}$ the orbital angular momentum, and $\eta=\pm 1$ the spin of the electron.
The corresponding eigenenergies are
\begin{align}\label{s:FD}
E_{n,l,\eta}^{(i)} = {}&{} \hbar \Omega (2 n + |l| + 1) - \frac{1}{2} \hbar \omega_c l + \frac{1}{4}g\frac{m^*}{m_e}\hbar\omega_c\eta,
\end{align}
where $m_e$ is the bare electron rest mass.
The first term contributes the regular two-dimensional harmonic-oscillator energies, but with a magnetic-field-dependent oscillator frequency $\Omega$, the second term adds the direct coupling of the angular momentum $l$ to the perpendicular magnetic field, and the last term accounts for the Zeeman effect.
In Fig.~\ref{s:fd} we show this spectrum as a function of $\omega_c$, where the boxed labels indicate the quantum numbers $(n,l)$ of the lowest few states.
For clarity we omitted the Zeeman effect, i.e., we set $g=0$, meaning that all lines are still twofold degenerate.
The levels plotted in red are the ones we used to construct the many-particles states that formed the basis for our analytic and numerical calculations, see below.
\begin{figure}[t]
	\includegraphics[scale=0.75]{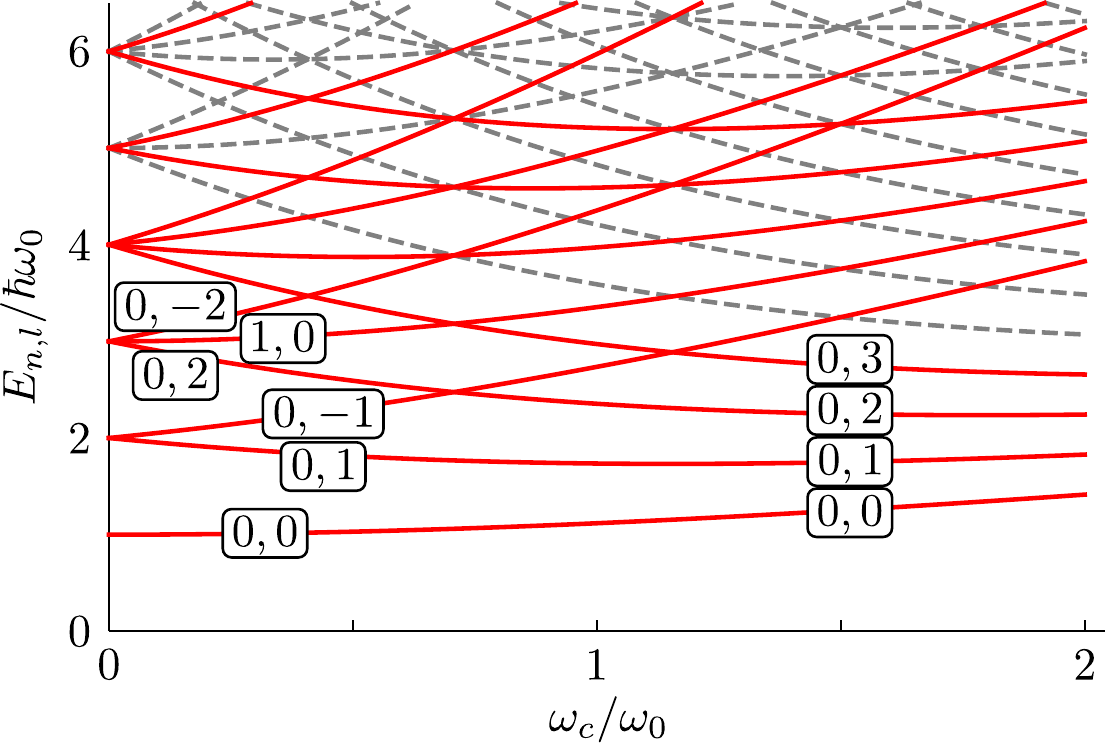}
	\caption{The Fock-Darwin spectrum as a function of $\omega_c = eB/m^*$ for $g=0$, see Eq.~(\ref{s:FD}), where the boxed labels indicate the orbital and angular momentum quantum numbers $(n,l)$.
	The levels plotted in red are the ones we included in constructing the many-particle states that we used as a basis for our calculations.}
	\label{s:fd}
\end{figure}

\subsection*{Many-particle Hamiltonian including interactions}

In a quantum dot with more than one electron one has to account for electron-electron interactions as well, which we describe by the Hamiltonian
\begin{align}\label{s:coulomb}
V = {}&{} \sum_{i<j} \frac{e^2}{4 \pi \varepsilon |\mathbf{r}_i - \mathbf{r}_j|},
\end{align}
where $\varepsilon$ is the effective dielectric constant of the surroundings of the quantum dot.
For a system with two electrons analytical diagonalization of the Hamiltonian $\sum_i H_0^{(i)} + V$ is possible~\cite{SGolovach2008}, but for more than two electrons there is no obvious solution.
We thus treat this many-body problem by working in a restricted configuration space, where we construct a basis of many-particle states from products of single-particle states and impose a cutoff on the quantum numbers $n$ and $l$ as was done in Ref.~\cite{SEto1997}.
In this basis, we can write explicit expressions for the matrix elements of the interaction Hamiltonian (\ref{s:coulomb}).
We then proceed by (\textit{i}) numerical diagonalization of the resulting Hamiltonian matrix and (\textit{ii}) applying perturbation theory in the interaction Hamiltonian $V$, which works best in the weak-coupling limit, where $e^2/4 \pi \varepsilon \sigma_0 \ll \hbar \omega_0$~\cite{SKouwenhoven1997}.

We start by constructing a basis of many-particle states from antisymmetrized products of Fock-Darwin states.
For a system of $M$ electrons one such product state, which we denote $\ket{s}$, is characterized by a set of quantum numbers $s=\{ n_{s_1}, l_{s_1},\eta_{s_1}; n_{s_2}, l_{s_2},\eta_{s_2}; \dots ; n_{s_M}, l_{s_M},\eta_{s_M} \}$.
The antisymmetrized wave function in position space $\langle \mathbf r_1, \mathbf r_2,\dots \mathbf r_M    |   s \rangle = \phi_s(\mathbf r_1,\mathbf r_2,\dots \mathbf r_M)$ can then be written as
\begin{align}
\phi_s(\mathbf r_1,\mathbf r_2,\dots \mathbf r_M) = \mathcal A [\psi_{n_{s_1},l_{s_1},\eta_{s_1}} (\mathbf r_1) \psi_{n_{s_2},l_{s_2},\eta_{s_2}} (\mathbf r_2) \dots \psi_{n_{s_M},l_{s_M},\eta_{s_M}} (\mathbf r_M)],
\end{align}
where $\mathcal A$ is the antisymmetrization operator.

Since $\sum_i H_0^{(i)}$ is diagonal in the basis of these product states, we can write the full Hamiltonian as
\begin{align}\label{s:h1}
H = {}&{} \sum_s \left[\hbar \Omega (2 N_s + K_s + M) - \frac{1}{2} \hbar \omega_c L_s + \frac{1}{2}g\frac{m^*}{m_e}\hbar\omega_c S_s + V_{ss}  \right] \ket s \bra s + \sum_{s\neq r} V_{sr} \ket s \bra r,
\end{align}
where $N_s = \sum_i n_{s_i}$, $K_s = \sum_i |l_{s_i}|$, $L_s = \sum_i l_{s_i}$, $S_s = \frac{1}{2} \sum_i \eta_{s_i}$, and $V_{sr}= \bmk{s}{V}{r}$.
To write an explicit matrix form of this Hamiltonian we therefore need to evaluate the integrals
\begin{align}
V_{sr} = {}&{} \sum_{i<j}^M \int d\mathbf r_1 \cdots d\mathbf r_M\, \phi_s^*(\mathbf r_1,\dots \mathbf r_M) \frac{e^2}{4 \pi \varepsilon |\mathbf{r}_i - \mathbf{r}_j|} \phi_r(\mathbf r_1,\dots \mathbf r_M)
\end{align}
for all sets of quantum numbers $s$ and $r$. Since the Coulomb potential couples electrons pairwise, we only need to evaluate integrals of the form
\begin{align}
v_{n_1,l_1;n_2,l_2;n_3,l_3;n_4,l_4} \equiv \int d\mathbf r_1 d\mathbf r_2 \,
\psi_{n_1,l_1,\eta}^* (\mathbf r_1) \psi_{n_2,l_2,\eta'}^* (\mathbf r_2) \psi_{n_3,l_3,\eta'} (\mathbf r_2) \psi_{n_4,l_4,\eta} (\mathbf r_1)
\frac{e^2}{4 \pi \varepsilon |\mathbf{r}_1 - \mathbf{r}_2|}.
\end{align}
Using the Fock-Darwin states as given in Eq.~(\ref{s:fdstates}) the result of the integral can be written in a closed form~\cite{SRontani1999a},
\begin{align}
v_{n_1,l_1;n_2,l_2;n_3,l_3;n_4,l_4} = {} & {}
\frac{e^2}{4\sqrt 2 \pi\varepsilon\sigma} \delta_{l_1+l_2,l_3+l_4} \sqrt{ \prod_{i=1}^4 \frac{n_i!}{2^{|l_i|}(n_i+|l_i|)!}}
\sum_{j_1=0}^{n_1}\sum_{j_2=0}^{n_2}\sum_{j_3=0}^{n_3}\sum_{j_4=0}^{n_4}
\left[ \prod_{k=1}^4 \frac{(-\frac{1}{2})^{j_k}}{j_k!} \left( \begin{array}{c} n_k+|l_k| \\ n_k - j_k \end{array} \right) \right]
\nonumber\\ {} & {}
\times \sum_{\lambda_1 = 0}^{\alpha_{1}}\sum_{\lambda_2 = 0}^{\alpha_{2}}\sum_{\lambda_3 = 0}^{\alpha_{3}}\sum_{\lambda_4 = 0}^{\alpha_{4}} \delta_{\lambda_1+\lambda_2,\lambda_3+\lambda_4} \left[ \prod_{t=1}^4 \left(\begin{array}{c} \alpha_{t} \\ \lambda_t \end{array}\right)\right]
(-1)^{\alpha_2+\alpha_3-\lambda_2-\lambda_3}\,
\Gamma \left( \frac{\Lambda+2}{2} \right)\Gamma\left( \frac{A-\Lambda +1}{2} \right),
\end{align}
where $\alpha_{i} = j_i+j_{5-i}+\frac{1}{2}(|l_i|+l_i+|l_{5-i}|-l_{5-i})$, $\Lambda = \sum_{i=1}^4 \lambda_i$, and $A = \sum_{i=1}^4 \alpha_i$.
This allows us to find analytic expressions for all $V_{sr}$ in (\ref{s:h1}) and thus to write $H$ in a closed matrix form.

\subsection*{Numerical results}

We will investigate many-particle states with up to 5 electrons in a single quantum dot, assuming relatively small applied magnetic fields, $\omega_c/\omega_0 \leq 0.3$, and not too strong interactions, characterized by the dimensionless parameter $\kappa = e^2/4\pi\varepsilon\sigma_0 \hbar\omega_0 \lesssim 1$.
Since we are only interested in finding the lowest few levels for $M \leq 5$, the 28 single-particle states with $n\in \{ 0,1 \}$, $l\in \{ 0,\pm 1,\pm 2,\pm 3 \}$ and $\eta = \pm 1$ (the levels plotted in red in Fig.~\ref{s:fd}, which all still have a twofold spin degeneracy) form in this case a reasonable set to construct our basis of many-particle states from.

\begin{figure}[t]
	\includegraphics[scale=0.63]{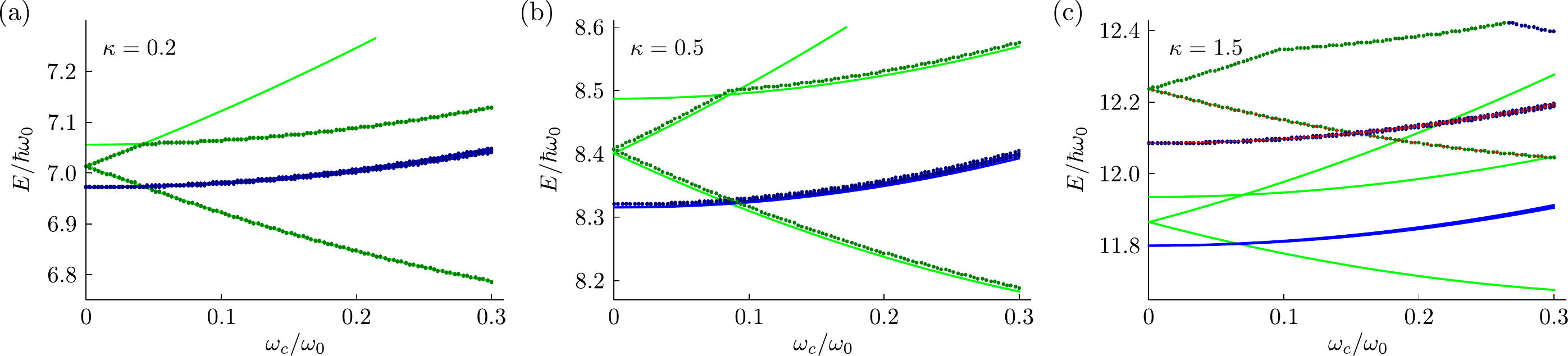}
	\caption{The lowest few four-electron levels in a single planar quantum dot as a function of the applied perpendicular magnetic field $\omega_c = eB/m^*$ with $g=-0.4$, for three different interaction parameters: (a) $\kappa=0.2$, (b) $\kappa=0.5$, and (c) $\kappa=1.5$.
	The dots are the result of numerical diagonalization of the Hamiltonian (\ref{s:h1}) written in the basis of antisymmetrized product states of the 28 single-particle states with $n\in \{ 0,1 \}$, $l\in \{ 0,\pm 1,\pm 2,\pm 3 \}$, and $\eta = \pm 1$.
	States that have a singlet spin configuration are plotted in green and states with a triplet spin configuration are plotted in blue.
	The solid lines show the analytic expressions for the field-dependent energies (\ref{s:eigsT}--\ref{s:gammas2}) as found from second-order perturbation theory in $\kappa$.
	The thin red dashed lines in (c) show the result of fitting the numerical data to Eqs.~(\ref{s:fit1}) and (\ref{s:fit2}), which have the same algebraic structure as the perturbative results given in (\ref{s:eigsT}) and (\ref{s:eigsS1}).}
	\label{s:spec}
\end{figure}
We now focus on the case of four electrons and proceed by diagonalizing the Hamiltonian numerically to obtain the eigenstates and -energies of the lowest electronic states in the multi-electron quantum dot.
The results are shown as dots in Fig.~\ref{s:spec}, where we used (a) $\kappa = 0.2$, (b) $\kappa=0.5$, and (c) $\kappa = 1.5$, and we set $g=-0.4$.
Green dots correspond to eigenstates that have a many-particle spin-singlet structure and blue dots to states with a spin-triplet structure.
Under the simplest assumptions that the surroundings of the dot are made of pure GaAs and that there are no additional screening effects, we use $\varepsilon = 12.9\,\varepsilon_0$ and $m^* = 0.067\,m_e$ to estimate the dot size and orbital level splitting corresponding to these two values of $\kappa$, giving $E_{\rm orb} \approx 40~$meV for $\kappa=0.5$ and $E_{\rm orb} \approx 5~$meV for $\kappa=1.5$.
For the most common gate-defined dots $\kappa = 1.5$ seems thus to be more realistic, although we note that the actual value of $\varepsilon$ is probably hard to predict since it can be affected severely by structural inhomogeneities, screening effects due to nearby metallic gates, or the underlying three-dimensional nature of the electronic wave function~\cite{SZhang1986}.

The overall structure of the low-field part of the spectra shown in Fig.~\ref{s:spec} is, however, the same for all values of $\kappa$:
At zero field, exchange effects arising from the Coulomb interaction favor a spin triplet ground state, which for small $\kappa$ has the orbital configuration $(0,0)^2(0,1)^1(0,-1)^1$, where the superscript denotes the number of electrons in the state $(n,l)$~\cite{SEto1997}.
The first excited states are two spin singlets with the configurations $(0,0)^2(0,1)^2$ and $(0,0)^2(0,-1)^2$, and the next excited state is the singlet with the configuration $(0,0)^2(0,1)^1(0,-1)^1$.
When the field is increased, the most pronounced effects are: (\textit{i}) all orbital energies increase due to the magnetic compression of the wave functions, i.e., the dependence of $\Omega$ on $\omega_c$, (\textit{ii}) the first two excited singlets split in energy due to their total projected orbital angular momentum of $L_z = \pm 2\hbar$, and (\textit{iii}) the three triplet states split due to the Zeeman effect.

All together, this leads to a singlet-triplet crossing at $\omega_c/\omega_0 \sim 0.1$ after which a four-particle singlet becomes the ground state.
Close to this degeneracy the next excited state is typically $\sim 0.2\,\hbar\omega_0$ higher in energy, which is far enough to treat the lowest four levels to first approximation as a well-separated subsystem.
One could use levels that cross in this subsystem to construct a singlet-triplet qubit~\cite{SJacak2001a}, but in this case qubit control would still rely on modulation of the magnetic field.
If, instead, we add two more quantum dots with a single electron on each, then we can create a triple-dot exchange-only singlet-only qubit similar to the quadruple-dot qubit proposed in Ref.~\onlinecite{SSala2017}, where the tunability of the singlet-triplet splitting of the two central electrons, through the detuning and coupling between the two central dots, is now replaced by tunability of the splitting through the external magnetic field.

\vspace{6ex}
\subsection*{Analytic results}

Since we have closed-form expressions for all elements of the interaction Hamiltonian, we can do perturbation theory in small $\kappa$ to arrive at analytic expressions for the lowest few eigenenergies in a multi-electron quantum dot~\cite{SRontani1999}.
For the four-electrons case studied above we find for the three lowest triplet levels
\begin{align}\label{s:eigsT}
E_{T_\beta^{(S)}} = 6\hbar\Omega + \frac{S}{2}g\frac{m^*}{m_e}\hbar\omega_c + \gamma^{(1)}_T \kappa \hbar\sqrt{\Omega\omega_0} + \gamma_T^{(2)}\kappa^2\hbar\omega_0,
\end{align}
where $S \in \{-1,0,1\}$ labels the total spin projection of the triplet and we used the coefficients
\begin{align}
	\gamma_T^{(1)} = 2\sqrt{2 \pi}, \qquad
	\gamma_T^{(2)} = -\frac{195893509 \pi}{805306368} \approx -0.764,
\end{align}
that determine the prefactor of the first- and second-order correction, respectively.
The two lowest singlet levels have
\begin{align}\label{s:eigsS1}
	E_{S_{\alpha,\gamma}} =  6\hbar\Omega - \frac{L}{2}\hbar\omega_c + \gamma^{(1)}_{S1} \kappa \hbar\sqrt{\Omega\omega_0} + \gamma_{S1}^{(2)}\kappa^2\hbar\omega_0,
\end{align}
where $L$ labels the total orbital angular momentum projection of the state, i.e., $L = 2$ for the lowest singlet $\ket{S_\alpha}$ and $L = -2$ for the first excited singlet $\ket{S_\gamma}$.
Further,
\begin{align}
	\gamma_{S1}^{(1)} = \frac{67}{16} \sqrt{\frac{\pi}{2} }, \qquad
	\gamma_{S1}^{(2)} = -\frac{38109479 \pi}{134217728} \approx -0.892.
\end{align}
Finally, for the singlet $\ket{S_\beta}$ that lives in the same combination of orbital states as the lowest triplet we find
\begin{align}\label{s:eigsS2}
	E_{S_\beta} =  6\hbar\Omega  + \gamma^{(1)}_{S2} \kappa \hbar\sqrt{\Omega\omega_0} + \gamma_{S2}^{(2)}\kappa^2\hbar\omega_0,
\end{align}
with
\begin{align}
	\gamma_{S2}^{(1)} = \frac{35}{8} \sqrt{\frac{\pi}{2}}, \qquad
	\gamma_{S2}^{(2)} = -\frac{1391260025 \pi}{4294967296} \approx -1.02.\label{s:gammas2}
\end{align}
These results are shown as solid lines in Fig.~\ref{s:spec}(a,b) and match the numerical data reasonably well---better for smaller values of $\kappa$---but for $\kappa = 1.5$ the perturbative results are off by $\sim$~5\%.
Using these results, we can also easily write down an expression for the singlet-triplet splitting denoted $E_{ST}$ in the main text,
\begin{align}
	 E_{T_\beta^0} - E_{S_\alpha} = \hbar\omega_c - \frac{3}{16} \sqrt{\frac{\pi}{2}} \kappa \hbar\sqrt{\Omega\omega_0} + \frac{32763365 \pi}{805306368} \kappa^2\hbar\omega_0,
\end{align}
which for small $\omega_c/\omega_0$ can be very well approximated by
\begin{align}
	E_{T_\beta^0} - E_{S_\alpha} \approx \hbar\omega_c  + \left( - \frac{3}{16} \sqrt\frac{\pi}{2} \kappa + \frac{32763365 \pi}{805306368} \kappa^2\right) \hbar\omega_0
	\approx \hbar\omega_c  + \left( -0.235\,\kappa + 0.128\,\kappa^2\right) \hbar\omega_0.
\end{align}

For larger $\kappa$, such as $\kappa = 1.5$, the level structure of the low-energy part of the spectrum still looks qualitatively very similar to the small-$\kappa$ case, which suggests that we can use the same type of algebraic expression to describe the energies.
If we write for the lowest four levels
\begin{align}
	E_{T_\beta^{(S)}} = {} & {} 6\hbar\Omega + \frac{S}{2}g\frac{m^*}{m_e}\hbar\omega_c + a_{1} \hbar\sqrt{\Omega\omega_0} + b_1 \hbar\omega_0,\label{s:fit1}\\
	E_{S_\alpha} =  {} & {} 6\hbar\Omega - \hbar\omega_c + a_2\hbar\sqrt{\Omega\omega_0} + b_2\hbar\omega_0,\label{s:fit2}
\end{align}
then we can use a least-square fit to the numerical data shown in Fig.~\ref{s:spec}(b) to extract the four parameters
\begin{align}
	a_1 = 7.383, \qquad b_1 = -1.146, \qquad
	a_2 = 7.123, \qquad b_2 = -1.037,
\end{align}
resulting in the red dashed curves shown in the Figure, which agree very well with the numerically calculated results.

\section*{Singlet basis states and Schrieffer-Wolff transformation}

As explained in the main text, the system is modeled using a Hubbard-like Hamiltonian,
\begin{equation}\label{s:hubbard}
{H} = \sum_i^3\left( H_1^{(i)} - V_i{n}_i\right) + \sum_{\langle i,j\rangle}U_c{n}_i{n}_j - \sum_{\langle i,j\rangle,\eta}\frac{t_{ij}}{\sqrt{2}}{c}_{i,\eta}^\dagger{c}_{j,\eta},
\end{equation}
where  $H_1^{(i)}$ is the many-particle Hamiltonian in Eq.~(\ref{s:h1}) acting on the electrons in dot $i$,
the $V_i$ describe the gate-tunable offset voltages on the three dots,
${n}_i=\sum_\eta{c}_{i,\eta}^\dagger{c}_{i,\eta}$ is the electron number operator for dot $i$, 
$U_c$ characterizes the interdot electrostatic coupling between the electrons,
the $t_{ij}$ describe tunneling between neighboring dots $i$ and $j$, and ${c}_{i,\eta}$ annihilates an electron on dot $i$ with spin $\eta$.

We assume six electrons occupying a linear array of three quantum dots.
Our qubit is defined in the singlet subspace of the (1,4,1) charge configuration, the lowest four singlet states being
\begin{align}
	\ket 0 = {}&{} \ket{S_{\alpha} S_{(13)}}, 
	\label{s:ket0}
	\\
	\ket 1 = {}&{} \frac{1}{\sqrt 3}\left[\ket{T_{\beta}^{0} T_{(13)}^{0}} - \ket{T_{\beta}^{-} T_{(13)}^{+}} - \ket{T_{\beta}^{+} T_{(13)}^{-}} \right],
	\label{s:ket1}
	\\
	\ket 2 = {}&{} \ket{S_{\beta} S_{(13)}}, 
	\label{s:ket2}
	\\
	\ket 3 = {}&{} \ket{S_{\gamma} S_{(13)}}, 
	\label{s:ket3}
\end{align}
where $\ket{S_{(13)}}$ and $\ket{T_{(13)}}$ indicate pairing in a singlet or triplet state of the two electrons in the outer dots, and $\ket{S_{\alpha,\beta,\gamma}}$ and $\ket{T_{\beta}}$ are the lowest four-particle singlets and triplet on the central dot, see the main text and above.
In the absence of interdot tunneling, i.e., for $t_{12}=t_{23}=0$, the energies of these four states are
\begin{align}
	E_0 = {} & {} 2E_{D_0^{(1,1)}} + E_{S_\alpha^{(4,2)}} - V_1 - 4V_2 - V_3 + 8U_c, \\
	E_1 = {} & {} 2E_{D_0^{(1,1)}} + E_{T_\beta^{(4,2)}} - V_1 - 4V_2 - V_3 + 8U_c, \\
	E_2 = {} & {} 2E_{D_0^{(1,1)}} + E_{S_\beta^{(4,2)}} - V_1 - 4V_2 - V_3 + 8U_c, \\
	E_3 = {} & {} 2E_{D_0^{(1,1)}} + E_{S_\gamma^{(4,2)}} - V_1 - 4V_2 - V_3 + 8U_c.
\end{align}
Here, $E_{D_0^{(n,i)}}$ denotes the lowest doublet eigenenergy of the many-particle Hamiltonian (\ref{s:h1}) for the case of $n$ electrons on dot $i$, where we assumed that dots 1 and 3 are identical.
Note that these energies do not include the Zeeman energy, i.e., they represent the case $E_{\rm Z} = 0$~\cite{SNote1}.
Since the corresponding state $\ket{D_0^{(1,1)}}$ has the exact orbital configuration $(0,0)^1$ this means that we simply have $E_{D_0^{(1,1)}} = \hbar\Omega^{(1)}$, where the superscript (1) indicates that we have to use the orbital energy of dot 1 (we assume the lateral dots to have a slightly smaller size than the central one).
Similarly, $E_{S_\alpha^{(4,2)}}$ is the eigenenergy of the lowest four-particle singlet state on the central dot, such as investigated above, etc.

We would now like to introduce the effect of the tunnel coupling between the dots, to leading order in the coupling parameters $t_{l,r}$ denoting the tunneling coupling between the leftmost and rightmost two dots, respectively.
For that purpose, we need to consider virtual transitions to the neighboring charge configurations (1,3,2), (2,3,1), (1,5,0), and (0,5,1).
(The charge states (2,2,2) and (0,6,0) also border the (1,4,1) region, but a transition to one of these requires \emph{two} tunneling events.)
The states that are directly coupled to the four basis states (\ref{s:ket0}--\ref{s:ket3}) are
\begin{align}
    \ket{4} = {} & {} \ket{\{ D_0^{(1,1)} D_0^{(3,2)}\}_S S_0^{(2,3)} }, \\
    \ket{5} = {} & {} \ket{\{ D_0^{(1,1)} D_1^{(3,2)} \}_S S_0^{(2,3)}}, \\
    \ket{6} = {} & {} \ket{\{ D_0^{(1,1)} D_0^{(5,2)} \}_S}, \\
    \ket{7} = {} & {} \ket{\{ D_0^{(1,1)} D_1^{(5,2)} \}_S},
\end{align}
where we used the same notation as above, i.e., $\ket{S_0^{(2,3)}}$ denotes the ground state singlet formed by two electrons in dot 3, in addition to which we used $\ket{\{D_aD_b\}_S}$ to denote the spin singlet formed by the two doublets $\ket{D_a}$ and $\ket{D_b}$.
We see that states $\ket{4}$ and $\ket{5}$ are (1,3,2) states and $\ket{6}$ and $\ket{7}$ (1,5,0) states, and there are thus four more states, $\ket{8}$--$\ket{11}$, that are exactly the same but with the dot indices 1 and 3 interchanged.
The three- and five-particle doublet ground states $\ket{D_0^{(3,2)}}$ and $\ket{D_0^{(5,2)}}$ have a dominating orbital configuration of $(0,0)^2(0,1)^1$ and $(0,0)^2(0,1)^2(0,-1)^1$, respectively, and since the splitting to the first excited doublet states with main configurations $(0,0)^2(0,-1)^1$ and $(0,0)^2(0,1)^1(0,-1)^2$ is relatively small we included them in the perturbation theory; these first excited doublets are indicated with a subscript 1.
Using the same notation as before, the energies of these virtual states follow straightforwardly as
\begin{align}
	E_4 = {} & {} E_{D_0^{(1,1)}} + E_{D_0^{(3,2)}} + E_{S_0^{(2,3)}} - V_1 - 3V_2 - 2V_3 + 9U_c, \\
	E_5 = {} & {} E_{D_0^{(1,1)}} + E_{D_1^{(3,2)}} + E_{S_0^{(2,3)}} - V_1 - 3V_2 - 2V_3 + 9U_c, \\
	E_6 = {} & {} E_{D_0^{(1,1)}} + E_{D_0^{(5,2)}} - V_1 -5 V_2 + 5U_c, \\
	E_7 = {} & {} E_{D_0^{(1,1)}} + E_{D_1^{(5,2)}} - V_1 - 5 V_2 + 5U_c,
\end{align}
and the energies $E_{8,9,10,11}$ again by interchanging the dot indices 1 and 3.

We now include the tunnel couplings $t_{12} \equiv t_l$ and $t_{23} \equiv t_r$, and assuming that we are deep enough in the (1,4,1) region so that the energy differences to the other four charge states is much larger than the tunnel couplings we can evaluate the exchange effects perturbatively in $t_{l,r}$.
This is done using a Schrieffer-Wolff transformation, and gives up to second order in $t_{l,r}$ the effective (1,4,1) Hamiltonian
\begin{equation}\label{s:sw}
    {H}^{(1,4,1)} = \left(\begin{array}{cccc}
        0 & J_x & J_{0,2} & 0 \\
        J_x & E_{ST} + J_1 - J_0 & J_{1,2} & J_{1,3} \\
        J_{2,0} & J_{2,1} & E_{S_\beta^{(4,2)}} - E_{S_\alpha^{(4,2)}}+J_2-J_0 & J_{2,3} \\
        0 & J_{3,1} & J_{3,2} & E_{S_\gamma^{(4,2)}} - E_{S_\alpha^{(4,2)}}+J_3-J_0
    \end{array}\right),
\end{equation}
where we subtracted $E_0+J_0$ as a constant and defined $E_{ST} \equiv E_{T_\beta^{(4,2)}} - E_{S_\alpha^{(4,2)}}$.
The qubit regime we consider is where $(E_{S_{\beta,\gamma}^{(4,2)}} - E_{S_\alpha^{(4,2)}}) \gg E_{ST}$: the magnetic field is tuned not too far from the $S_\alpha T^0_\beta$-crossing in the four-electron central dot, so that the splitting between $\ket{S_\alpha}$ and $\ket{T_\beta^0}$ ($\sim 10~\mu$eV) is much smaller than the distance to the other two singlets $\ket{S_\beta}$ and $\ket{S_\gamma}$ ($\sim 0.5~$meV).
Since typical exchange energies are $J \sim 1~\mu$eV, we can, to first approximation, neglect the exchange-induced coupling of the qubit to the states $\ket{2}$ and $\ket{3}$; they would lead to small corrections of the order $\sim J^2 / (E_{S_{\beta,\gamma}^{(4,2)}} - E_{S_\alpha^{(4,2)}})$.
A second (more practical) reason to neglect these couplings is that these corrections are $\propto J^2 \propto t^4$, and for consistency one would then also have to perform the original Schrieffer-Wolff transformation to order $t^4$, now including the charge states (2,2,2) and (0,6,0) as well.

We then arrive at the effective qubit Hamiltonian
\begin{align}\label{s:hqb}
	H_\text{qb} = \frac{1}{2}(E_{ST} + J_z)\sigma_z + J_x \sigma_x,
\end{align}
where
\begin{align}
	J_z = {} & {} \ \frac{3t_r^2}{4}\left( \frac{1}{E_1-E_4} + \frac{1}{E_1-E_5}+ \frac{1}{E_1-E_6} + \frac{1}{E_1-E_{7}} \right)
	\nonumber\\&
	+ \frac{3t_l^2}{4}\left( \frac{1}{E_1-E_8} + \frac{1}{E_1-E_9}+ \frac{1}{E_1-E_{10}} + \frac{1}{E_1-E_{11}} \right)\nonumber\\
	{} & {} -\frac{t_r^2}{2} \left( \frac{1}{E_0-E_4} + \frac{1}{E_0-E_6} \right) - \frac{t_l^2}{2}\left( \frac{1}{E_0-E_8} + \frac{1}{E_0-E_{10}} \right),\\
	J_x = {} & {} \frac{\sqrt{3}t_r^2}{4\sqrt{2}}\left( \frac{1}{E_0-E_6} + \frac{1}{E_0-E_4}+ \frac{1}{E_1-E_6} + \frac{1}{E_1-E_{4}} \right)
	\nonumber\\&
	- \frac{\sqrt{3}t_l^2}{4\sqrt{2}}\left( \frac{1}{E_0-E_{8}} + \frac{1}{E_0-E_{10}}+ \frac{1}{E_1-E_{8}} + \frac{1}{E_1-E_{10}} \right).
\end{align}

The largest term in the Hamiltonian (\ref{s:hqb}) is
\begin{align}
	E_{ST} = \hbar\omega_c + (a_1-a_2)\hbar\sqrt{\Omega^{(2)}\omega^{(2)}_0} +(b_1-b_2)\hbar\omega^{(2)}_0,
\end{align}
in terms of the notation of Eqs.~(\ref{s:fit1}--\ref{s:fit2}), where $\Omega^{(2)}$ and $\omega_0^{(2)}$ are the oscillator frequencies (with and without magnetic field) of the central dot.
Here one can use $a_1 = \kappa\gamma_T^{(1)}$, $b_1 = \kappa^2\gamma_T^{(2)}$, $a_2 = \kappa\gamma_{S1}^{(1)}$, and $b_2 = \kappa^2\gamma_{S1}^{(2)}$ for $\kappa \lesssim 0.5$.

The exchange terms are relatively small and their approximate magnitude can be related to the width (and ``height'') $2\Delta$ of the stable (1,4,1) region in terms of the tuning parameters $V_d=\frac{1}{2}\left(V_3-V_1\right)$ and $V_m=\frac{1}{2}\left(V_1+V_3\right)-V_2$, respectively.
We assume that we can neglect the difference between $E_0$ and $E_1$ compared to the splitting to the other eight states.
For a given tuning $(V_m,V_d)$ the splitting to the states $\ket 4$, $\ket 6$, $\ket 8$, and $\ket{10}$ then equals the distance to the corresponding excited charge state in the charge stability diagram.
The energies of the four states that involve one excited orbital state, $\ket 5$, $\ket 7$, $\ket 9$, and $\ket{11}$, are higher in energy by $\hbar\omega_c$.
Assuming that the stable (1,4,1) region is roughly symmetric in $V_d$ and $V_m$, we then arrive at the approximate expressions
\begin{align}
	E_{0,1} - E_4 = {} & {} -\Delta+V_d+V_m,\\
	E_{0,1} - E_5 = {} & {} -\Delta+V_d+V_m-\hbar\omega_c,\\
	E_{0,1} - E_6 = {} & {} -\Delta-V_d-V_m,\\
	E_{0,1} - E_7 = {} & {} -\Delta-V_d-V_m-\hbar\omega_c,\\
	E_{0,1} - E_8 = {} & {} -\Delta-V_d+V_m,\\
	E_{0,1} - E_9 = {} & {} -\Delta-V_d+V_m-\hbar\omega_c,\\
	E_{0,1} - E_{10} = {} & {} -\Delta+V_d-V_m,\\
	E_{0,1} - E_{11} = {} & {} -\Delta+V_d-V_m-\hbar\omega_c,
\end{align}
With these approximations we find
\begin{align}
	J_z = {} & {} \frac{1}{2}
	\left( \frac{t_l^2\Delta}{(V_d-V_m)^2-\Delta^2} 
	+ \frac{t_r^2\Delta}{(V_d+V_m)^2-\Delta^2}
	+  \frac{3t_l^2(\Delta+\omega_c)}{(V_d-V_m)^2-(\Delta+\omega_c)^2} 
	+ \frac{3t_r^2(\Delta + \omega_c)}{(V_d+V_m)^2-(\Delta+\omega_c)^2}
	\right), \\
	J_x = {} & {} \sqrt{\frac{3}{2}}
		\left( \frac{t_r^2\Delta}{(V_d+V_m)^2-\Delta^2} - \frac{t_l^2\Delta}{(V_d-V_m)^2-\Delta^2}\right).
\end{align}
Assuming approximately equal tunnel couplings $t_{l,r}$ and qubit operation near the ``line'' where $V_d = 0$, we define $t = \frac{1}{2}(t_l+t_r)$ and $\delta t = t_l - t_r$, and expand the exchange energies to leading order in $\delta t$ and $V_d$,
\begin{align}
J_z \approx {} & {} -t^2\left[ \frac{\Delta}{\Delta^2 - V_m^2} + \frac{3(\Delta + \hbar\omega_c)}{(\Delta + \hbar\omega_c)^2 - V_m^2} \right], \\
J_x \approx {} & {} \frac{\sqrt 6 t \Delta}{\Delta^2 - V_m^2} \left[ \delta t + \frac{2 t V_m}{\Delta^2 - V_m^2}V_d \right].
\end{align}

\vspace{4ex}
\section*{Higher-order hyperfine interaction and dephasing}
Hosting the qubit in singlet states only results in having no direct coupling between the qubit states via the hyperfine interaction to lowest order.
Higher-order effects, however, can give rise to energy shifts of the qubit splitting that may lead to qubit dephasing. 

We treat the hyperfine interaction between the electron spins and the spins of the many nuclei on a mean-field level, resulting in a Zeeman-like Hamiltonian
\begin{equation}
{H}_\text{hf} = \frac{g\mu_B}{2}\sum_{i}\mathbf{K}_i\cdot {\boldsymbol\sigma}_i,\label{s:hf}
\end{equation}
where $\mathbf{K}_i$ is the effective nuclear field acting on electron $i$ and $(\hbar/2){\boldsymbol\sigma}_i$ is the spin operator for electron $i$.
Due to the tiny nuclear magnetic moments, the nuclear-spin density matrix will be in a high-temperature mixed state for all experimentally relevant temperatures.
This results in random nuclear fields ${\bf K}_i$ that have zero mean and a standard deviation $\sigma_K \sim A/\sqrt N$, where $A$ is an effective material-dependent hyperfine coupling parameter and $N$ the number of nuclear spins the electron is coupled to.
For typical GaAs-based quantum dots $\sigma_K \sim 1$--5~mT.

The main contribution to higher-order hyperfine terms in the qubit subspace comes from other (non-qubit) spin states that are close in energy.
There are three triplets that are energetically close to the qubit states (labeled $\ket{T_{1,2,3}^0}$ in Fig.~3 in the main text), one lying below $\ket 0$ and two in between $\ket 1$ and the quintuplet $\ket Q$.
The energy splitting between the qubit states and the triplets are governed by exchange effects, where the splitting between $\ket{0}$ and $\ket{T_1^0}$ goes as $\sim t^4/\Delta^3$ and the splitting between $\ket{1}$ and $\ket{T_2^0}$ and $\ket{T_3^0}$ as $\sim t^2/\Delta$.
Thus, the hyperfine-induced shift in the qubit splitting is dominated by the coupling between $\ket{0}$ and the triplet states 
\begin{align}
	\ket{T_1^{+}} = {} & {} \ket{S_\alpha T^{+}_{(13)}}, \\
	\ket{T_1^{0}} = {} & {} \ket{S_\alpha T^{0}_{(13)}}, \\
	\ket{T_1^{-}} = {} & {} \ket{S_\alpha T^{-}_{(13)}}.
\end{align}

We project the hyperfine Hamiltonian (\ref{s:hf}) to the subspace $\{\ket{0},\ket{T_1^{+}},\ket{T_1^{0}},\ket{T_1^{-}}\}$, yielding
\begin{equation}
{H}_\text{hf}
= \left(\begin{array}{cccc}
0 & \iota^-_{13} & \kappa_{13} & \iota^+_{13} \\
\iota^+_{13} & 0 & 0 & 0 \\
\kappa_{13} & 0 & 0 & 0 \\
\iota^-_{13} & 0 & 0 & 0 
\end{array}\right),
\end{equation}
where we defined the gradients
\begin{align}
	\kappa_{13} = {} & {} \frac{g\mu_B}{2}\left(K^z_1 - K^z_3\right), \\
	\iota_{13}^\pm = {} & {} \frac{g\mu_B}{2\sqrt{2}}\left(K^\pm_1 - K^\pm_3\right)
	= \frac{g\mu_B}{2\sqrt{2}}\left(K^x_1 \pm iK^y_1 - K^x_3 \mp iK^y_3\right).
\end{align}
Using perturbation theory, we find that, to leading order in $K_i^{x,y,z}$ the energy shift of $\ket{0}$ is 
\begin{align}
	\delta E_0 = {} & {} \frac{\kappa_{13}^2}{E_0-E_{T_1^0}} + \frac{\iota_{13}^+\iota_{13}^-}{E_0-E_{T_1^{+}}} + \frac{\iota_{13}^+\iota_{13}^-}{E_0-E_{T_1^{-}}},
\end{align}
where the $E_\nu$ are the unperturbed energy levels.
Due to the larger separation between $\ket 1$ and $\ket{T_{2,3}^0}$, the hyperfine-induced shift of $\ket{1}$ is much smaller.
The shifts caused by the coupling to the polarized triplets $\ket{T_1^{\pm}}$, i.e.\ the two last terms, can be reduced by tuning the Zeeman energy, which can be done independently from $\omega_c$ by tilting the total externally applied field.
The energy of $\ket{T_1^0}$ is not affected by the Zeeman effect, and the shift caused by the coupling to this state is thus solely determined by exchange effects.

To obtain a very rough estimate for the scale of the dephasing time caused by these higher-order hyperfine fields, we consider their effect on the Rabi oscillations when the system is driven resonantly.
Following the approach of Ref.~\cite{SSala2017} we find the estimate $T_2^* \sim A_q\hbar(E_0-E_{T_1^0})^2/\sigma_{K}^4$, which indeed predicts a shorter dephasing time the closer the state $\ket{T_1^0}$ is to $\ket{0}$, i.e., a shorter hyperfine-induced $T_2^*$ at the sweet spot than in the resonant-exchange regime.

To illustrate, we calculate numerically the probability $\langle P_1(\tau)\rangle$ of finding the qubit in $\ket 1$ after initializing in $\ket 0$ and resonantly driving either $V_d$ or $\delta t$ for a time $\tau$, focusing on the two cases illustrated in Fig.~3(a,b) in the main text.
In Fig.~\ref{fig:supp:dephasing} we show the resulting time-dependent probabilities for (a) driving $V_d$ in the resonant-exchange regime, and (b) driving $\delta t$ at the sweet spot, after averaging over 2500 random nuclear configurations with $g\mu_B K_i^{x,y,z}$ each taken from a normal distribution with mean zero and $\sigma_{K} = 0.07\ \mu$eV (corresponding to 3~mT for $|g| = 0.4$).
We used the same parameters as in the main text, with driving amplitudes $\tilde V_d=10\ \mu$eV and $\delta\tilde t=2\ \mu$eV.
The results show that the hyperfine-induced dephasing, even at the sweet spot, is small compared to the Rabi period.

We finally compare this with the rough estimate $T_2^* \sim A_q\hbar(E_0-E_{T_1^0})^2/\sigma_{K}^4$.
Using the fact that the Rabi period is given by $T_{\rm Rabi} = h/A_q$, we find for the approximate number of coherent Rabi oscillations that should be visible
\begin{align}
	n_{\rm coh} \equiv \frac{T_2^*}{T_{\rm Rabi}} \sim \frac{1}{2\pi} \frac{A_q^2(E_0-E_{T_1^0})^2}{\sigma_{K}^4} = 
	\frac{1}{2\pi} \frac{h^2(E_0-E_{T_1^0})^2}{T_{\rm Rabi}^2\sigma_{K}^4}.
\end{align}
For the resonant-exchange regime of Fig.~\ref{fig:supp:dephasing}(a) we read off $T_{\rm Rabi}\approx 25$~ns and we find $E_0-E_{T_1^0} \approx 0.80~\mu$eV which yields $n_{\rm coh} \sim 120$, whereas at the sweet spot, see Fig.~\ref{fig:supp:dephasing}(b), we have $T_{\rm Rabi}\approx 19$~ns and $E_0-E_{T_1^0} \approx 0.32~\mu$eV giving $n_{\rm coh} \sim 30$.
We see that these estimates are indeed roughly agreeing with the dephasing observed in Fig.~\ref{fig:supp:dephasing}.

\begin{figure}
	\centering
	\includegraphics[width=.75\textwidth]{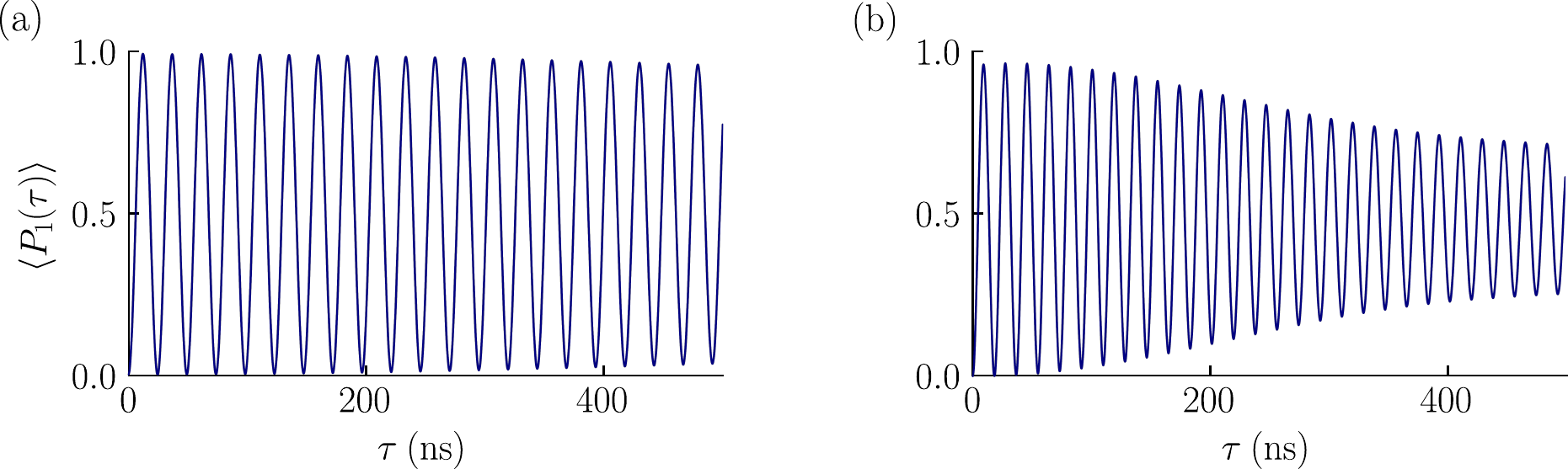}
	\caption{Time dependent probability $\langle P_1(\tau)\rangle$ after averaging over 2500 random nuclear field configurations, (a) in the resonant-exchange regime and (b) at the sweet-spot.
	We used the same parameters as in the main text, with $\tilde V_d = 10\ \mu$eV, $\delta\tilde t= 2\ \mu$eV, and $g\mu_B K_i^{x,y,z}$ taken from a normal distribution with zero mean and $\sigma_{K} = 0.07\ \mu$eV.}
	\label{fig:supp:dephasing}
\end{figure}

\end{NoHyper}


\begin{thebibliography}{39}
\bibitem [{foo()}]{footnote1}%
	\BibitemOpen
	\href@noop {} {}\bibinfo {note} {The dependence of the orbital spacing on the
		dot size allows us to design a device with an orbital splitting on the outer
		dots larger than the level splittings in the central multi-electron dot,
		which enables full spin blockade in the regime where the qubit is
		operated.}\BibitemShut {Stop}%
\end{thebibliography}

%

\end{document}